\documentclass[11pt]{article}
\usepackage{graphicx}

\begin{document}
%%%%%%%%%%%%%%%%%%%%%%%%%%%%%%%%%%%%%%%%%%%%%%%%%%%%%%%%%
\title {Quantum Criticality in Ferromagnetic Single-Electron Transistors}

\author{\hspace*{-1cm}\small Stefan Kirchner$^{1}$, Lijun Zhu$^{1,2}$, 
Qimiao Si$^{1,}$\footnote{To whom correspondence should be 
addressed. Email: qmsi@rice.edu.}\, \& D. Natelson$^{1}$\\[2mm]
{\hspace*{-1cm}\small $^1$Department of Physics \& Astronomy, Rice University, Houston,
TX 77251-1892, USA }\\
{\hspace*{-1cm}\small $^2$ Department of Physics, University of California, Riverside,
CA 92521, USA}}
\date{}
\maketitle
\begin{abstract}{
Considerable evidence exists for the failure
of the traditional theory of quantum critical points (QCPs),
pointing to the need to incorporate novel excitations. 
The destruction of Kondo entanglement and the concomitant
critical Kondo effect
may underlie these emergent excitations in
heavy fermion metals -- a prototype system for quantum criticality
--  but the effect remains poorly understood. Here, we show how
ferromagnetic single-electron transistors
can be used to study this effect. 
We theoretically demonstrate a gate-voltage induced quantum phase
transition. The critical Kondo effect is manifested in
a fractional-power-law dependence of the conductance on temperature
($T$). The AC conductance and thermal noise spectrum have related
power-law dependences on frequency ($\omega$) and, in addition,
show an $\omega/T$ scaling. Our results imply that the ferromagnetic
nanostructure constitutes a realistic model system
to elucidate magnetic quantum criticality
that is central to the heavy fermions and other bulk materials with
non-Fermi liquid behavior.
}
\end{abstract}

%%\newpage

A quantum critical point (QCP) occurs at zero temperature
as a system changes from one ground state to another, and 
controls physical properties over a wide region in the phase 
diagram at finite 
temperatures~\cite{Coleman.05,Si.01,Paschen.04,Sachdev,Varma.02,Hertz}.
Electrons in condensed matter are traditionally described
as a Fermi liquid, a collection of essentially independent
particles.
Near a QCP, however, electrons are coupled
to each other in a singular fashion; how such electron correlations
lead to non-Fermi liquid states is an open issue that is central
to a variety of strongly correlated systems, including high
temperature superconductors and heavy fermion metals~\cite{Varma.02}.
Single-electron devices 
play a unique role in the study
of correlated electronic states, in addition to their potential
application to quantum electronics and quantum information processing.
The Fermi liquid state
has been studied systematically in semiconductor
quantum dots and single-molecule 
transistors~\cite{Glazman.88,Ng.88,Goldhaber-Gordon.98,CronenwettetAl98Science,Schmid.98,Park.02,LiangSBLP02,YuN04_b}.
Here, the low-energy excitations of the
normal metal leads are electrons near their
respective Fermi energies. These itinerant electrons
are entangled with the magnetic moment 
localized on the dot, producing a Kondo resonance.
The manifestation of the Kondo resonance
in the conductance spectrum was
predicted~\cite{Glazman.88,Ng.88} and was subsequently observed 
in semiconductor quantum
dots~\cite{Goldhaber-Gordon.98,CronenwettetAl98Science,Schmid.98}
and in
single-molecule transistors~\cite{Park.02,LiangSBLP02,YuN04_b}.
Ingenious proposals and structures~\cite{Oreg.03,Pustilnik.04,Craig.04}
have been put forth 
to
model the non-Fermi 
liquid states associated with the multichannel Kondo 
systems
and the two-impurity Kondo 
effect~\cite{Varma.02}.
The experimental observation of the non-Fermi liquid states 
is, however, still lacking,
in part because such states of quantum-impurity models
are not robust, requiring
special symmetries that are difficult to realize~\cite{Varma.02}.

Recently, it has become possible to fabricate 
single-electron transistors
with 
leads made of ferromagnetic metals~\cite{Pasupathy.04}. 
Fig.~1a provides a schematic
illustration.
The dot, which denotes
a quantum-mechanical spin (local moment) of a molecule or
that of a semiconductor quantum dot in the Coulomb-blockade regime,
is also capacitively coupled to a nonmagnetic gate electrode.
Here, we propose this structure as a model system 
to study how magnetism interplays with the Kondo effect, leading to a QCP
with non-Fermi liquid behavior.
Our key observation is that the ferromagnetic leads contain not
only conduction 
electrons
but also spin waves -- 
collective low-energy excitations 
arising due to the spontaneously broken spin symmetry of a ferromagnet.
We are then led to a description in terms of a Bose-Fermi Kondo
model~\cite{Si.01,Smith3,Sengupta.00,Zhu.02,Zarand.02},
which couples the local moment to both a fermionic bath 
of conduction electrons and a bosonic one of spin waves.

\begin{figure}[h!]
\centerline{\includegraphics[width=0.65\linewidth]{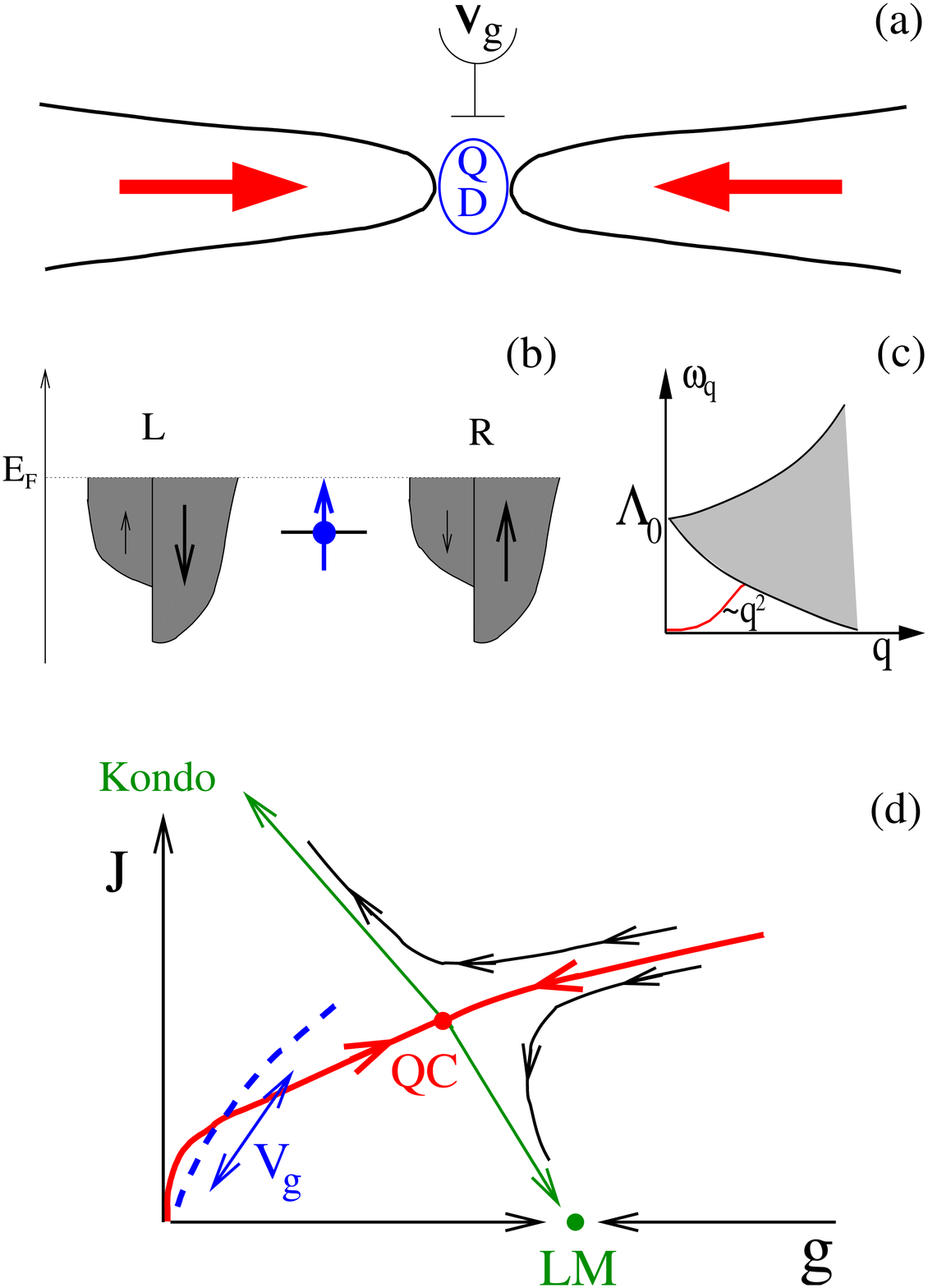}}
\caption{
Single-electron transistor with ferromagnetic leads.
{\bf a},~ A schematic set up. The opposing arrows denote antiparallel
(AP) alignment of the magnetizations of the ferromagnetic source and
drain. $V_g$ is the voltage of the normal-metallic gate. ``QD'' labels
the quantum-dot island of a semiconductor quantum dot or a molecule 
of a single-molecular transistor.
{\bf b},~The majority and minority electron bands of the left (``L'')
and right (``R'') leads. $E_F$ labels the Fermi energy. Also shown
is the local moment on the dot (blue arrow).
{\bf c},~Transverse spin excitations of the ferromagnetic leads.
The red line describes spin waves whose energy ($\omega_{\bf q}$)
depends on the wavevector (${\bf q}$) quadratically. The shaded area
denotes the continuum associated with particle-hole
excitations. The energy of the spin wave at the point it merges
into the continuum [$\Lambda$, not shown] is of the same order 
of magnitude as $\Lambda_0$, the energy cost to create a particle-hole
excitation at ${\bf q}=0$.
{\bf d},~The phase diagram of the Bose-Fermi Kondo model
[see Eq.~(\ref{hamiltonian-bfk-n=2}) of the Formal Details
section].
The parameters $J$ and $g$ are the couplings of the local
moment to the fermions and spin waves,
respectively. Lines with arrows denote the
RG flow. There are three RG fixed points. ``Kondo'' and ``LM''
refer to the Kondo screened fixed point, corresponding to 
a Fermi liquid, and the critical local-moment fixed point, 
describing a quantum-critical phase. ``QC'' refers to the
quantum-critical fixed point, characterizing the critical Kondo
screening on the entire separatrix (red line, corresponding to the 
critical coupling $g_c$ as a function of $J$).
Varying the gate voltage, $V_g$, tunes both $J$ and $g$,
along the blue-dashed line. 
} \label{FIG1}
\end{figure}

We focus on the 
antiparallel (AP) configuration of the magnetizations of the two 
ferromagnetic leads, and consider symmetric couplings between the dot and 
the left/right leads.  The fermionic bath ($c_{{\bf k}\sigma}$)
describes a linear combination of the conduction electron
bands of the two leads. The bosonic bath
contains two components, one each for the two orthogonal
spin 
polarization directions that are perpendicular to the
lead
magnetizations; each 
component
is a linear combination of the 
corresponding spin waves in the two leads.
We note on a number of features of the resulting Bose-Fermi Kondo
model
[see Eq.~(\ref{hamiltonian-bfk-n=2}) of the Formal Details
%%Methods 
section]
specific to the system.
First, while the energy bands of the
two spin components of the conduction electrons in each metallic
lead is (Zeeman) split by the magnetization, the linear combination
that forms the fermionic bath 
contains no Zeeman splitting
(Refs.~\cite{MartinekSBBKSv03,Choi.04}, and references therein)
as can be readily seen 
from Fig.~1b. Second,
in addition to fringing fields from the ferromagnetic leads, 
an effective local magnetic field ($h_{\rm loc}$) for the local moment
is usually generated due to 
dot-lead exchange coupling.
In our AP case, however, the contributions from the two leads
compensate with each other and $h_{\rm loc}$ vanishes.
Third, the two-component nature of the bosonic bath means that our model
has an easy-plane spin anisotropy. Finally, the dispersion of the
ferromagnetic spin waves ($\omega_{\bf q}$) is such that the spectral
density of the bosonic bath has a square-root dependence 
on the frequency,
\begin{eqnarray}
\sum_{\bf q}\delta (\omega - \omega_{\bf q}) \propto \sqrt{\omega} ,
\label{dispersion}
\end{eqnarray}
for $0<\omega<\Lambda$, the cutoff frequency below which spin waves
exist ({\it cf.} Fig.~\ref{FIG1}c).
The sub-linear dependence on frequency is 
commonly referred to
as 
``sub-ohmic'' dissipation.
We note that the gap of spin waves in real ferromagnets
is non-zero; its effect will be discussed shortly.

We find that a quantum phase transition can be induced by
varying the gate voltage, $V_g$.
This voltage determines $\Delta$, 
the valence-fluctuation energy cost of the dot:
$\Delta = \Delta_0 - c V_g$, where 
$\Delta_0$ is the corresponding energy cost in the absence 
of the gate voltage and the proportionality factor
$c = e C_{\rm gate}/C_{\rm tot}$ is determined by the
gate and total capacitances, $C_{\rm gate}$ and $C_{\rm tot}$,
respectively.
The parameter $\Delta$ in turn determines the strength of the 
interactions between the local moment and the bosonic and
fermionic baths. The local-moment regime arises when
$\Delta \gg \Gamma \equiv \pi \rho_0t^2$,
the non-interacting resonance width; here,
$t$ is the dot-lead hybridization 
matrix and $\rho_0$ the fermion density of states
at the chemical potential.
The resulting Bose-Fermi Kondo model is given
in Eq.~(\ref{hamiltonian-bfk-n=2}) of the Formal
Details section.
The Kondo exchange coupling with the fermions,
and the bare Kondo scale (at $g=0$),
have the standard forms: 
$J \sim \Gamma /\rho_0 \Delta$ and 
$T_K^0 \sim (1/\rho_0){\rm exp}(-\pi \Delta /\Gamma)$,
respectively. 
The coupling with the bosonic bath is 
found 
%%(see Methods section)
(see the Formal Details section)
to be
$g \sim \Gamma /(\rho_0\Delta)^2$.
Hence, the ratio $g / T_K^0$ is exponentially large,
and it decreases as $\Delta$ is reduced. 
When $\Delta$ becomes comparable to 
$\Gamma$,
the system reaches the mixed-valence 
regime,
and $g/T_K^0$ becomes of order unity.
Changing $V_g$ then takes the system through
second-order quantum phase transition
(Fig.~\ref{FIG1}d).

More specifically, for gate voltages such that the system (dot + leads)
is either in the mixed valence regime or in the local-moment regime but
with moderate ratios of $\Delta/\Gamma$, 
the usual Kondo effect takes place. At low temperatures,
the system is in a Fermi liquid ground state.
As $\Delta$ is increased by varying $V_g$, interactions with the spin
waves destroy the Kondo effect. Kondo resonances no longer
form,
and the system is in a non-Fermi liquid ground state.
In between the system passes through a QCP,
which is also a non-Fermi liquid.
These phases and the phase transition are expressed in terms of the
renormalization group (RG) flows and RG fixed points in 
Fig.~\ref{FIG1}d. At relatively small $g/T_K^0$, the RG flow
is towards the Kondo fixed point.
For sufficiently large $g$, the flow is instead towards 
a fixed point located on the $J=0$ axis (the horizontal line
of Fig.~\ref{FIG1}d) where the local moment is completely
decoupled from the fermions but is still strongly
fluctuating~\cite{Sachdev-Ye} through the coupling to the spin waves.
We will call this phase in the entire $g>g_c$ region  a
``critical local-moment phase''.
In between, the system passes through
$g=g_c$ where the RG flow is towards a quantum-critical
fixed point. Here, the local moment fluctuates in a critical fashion
(at time scales beyond the short-time cutoff, $\hbar /T_K^0$)
and the electronic excitations have a non-Fermi liquid
form.
The sub-ohmic nature of the dissipative bosonic bath 
[Eq.~(\ref{dispersion})] is necessary for 
the existence of quantum criticality.

%%\\[2ex]
\vspace*{0.4 cm}
\noindent{\bf \Large Results and Discussion}
%%\vspace*{0.2 cm}

\begin{figure}[h!]
\centerline{\includegraphics[width=1.0\linewidth]{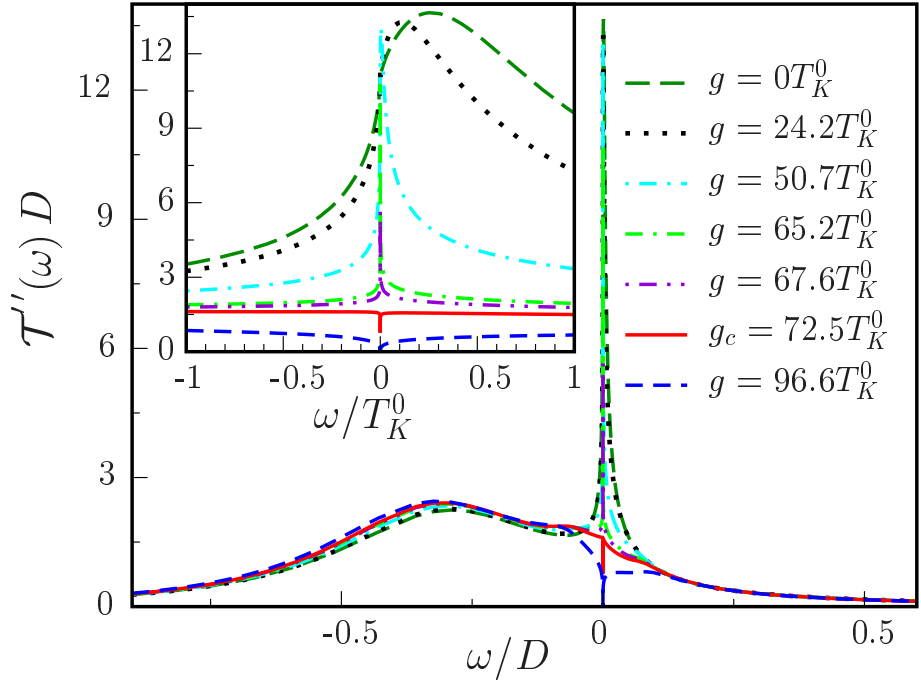}}
\caption{Evolution of the Kondo Resonance.
The scattering T-matrix, ${\cal T(\omega)}$, 
is defined as the Fourier transform of 
$-G_{d}(\tau) = <{\rm T}_{\tau} d_{\sigma} (\tau)
d_{\sigma}^{\dagger} (0)>$ 
(where $\tau$ is the imaginary time)
of the Bose-Fermi
Anderson model, 
${\cal T}''$ is the imaginary part of ${\cal T}$,
and is equal to $\pi$ times the single-electron
spectral function on the dot.
The main plot uses the fermionic half-bandwidth,
$D=1/2\rho_0$, as the normalization for energies.
The peak near $\omega=0$, arising for $g<g_c$, corresponds to the 
Kondo resonance.
The inset, limited to the energy range of about $T_K^0$,
highlights the vicinity of zero energy. 
Different $g/T_K^0$ corresponds to different
gate-voltage $V_g$. 
The parameters
adopted are:
$\epsilon_d=-0.3D$, $U=\infty$, $t=0.1D$, corresponding to 
$T_K^0=4.2 \times 10^{-3}D$;
the cut-off energy
for the bosonic bath 
[{\it cf.} Eq.~(\ref{dispersion})] is
$\Lambda = 0.05D$ and the proportionality factor 
on the right hand side of Eq.~(\ref{dispersion}) is
$1/\Gamma(3/2)$. 
The scaling exponents are universal,
but $g_c/T_K^0$ is not, depending on
$\rho_s$,
$\Lambda$ and other microscopic parameters 
of the ferromagnets. 
The same parameters apply to Figs.~(\ref{FIG3},\ref{FIG4}).
} \label{FIG2}
\end{figure}

To systematically study the transport properties across the quantum phase
transition, we apply a dynamical large-N 
approach~\cite{Zhu.04,Parcollet.98,Cox.93} to the Bose-Fermi 
Kondo model  
%%(see Methods section).
(see the Formal Details section).
We first consider the 
single-electron spectral function
on the dot, expressed in terms of the scattering T-matrix.
The results are shown in Fig.~\ref{FIG2}.
Since the universal properties are insensitive to the way
the quantum-critical regime is accessed, we have, for convenience,
chosen to work with fixed $T_K^0$ and varying $g$.
The different curves correspond to different ratios of $g/T_K^0$
and, correspondingly, different values of the gate-voltage $V_g$.
The spectral weight of the Kondo resonance is seen to decrease
as $g$ increases, and vanishes continuously when
$g$ reaches $g_c$.
At the QCP, the Kondo screening is critical: 
a Kondo resonance is no longer fully developed but vestiges of it 
remain.
For $g>g_c$, the Kondo effect is completely destroyed: the spectral density
vanishes at $\omega=0$. At finite energies, however, the spectral density
is still finite reflecting on a finite effective (scale-dependent, 
{\it or} renormalized) Kondo coupling.

\begin{figure}[h!]
\centerline{\includegraphics[width=1.0\linewidth]{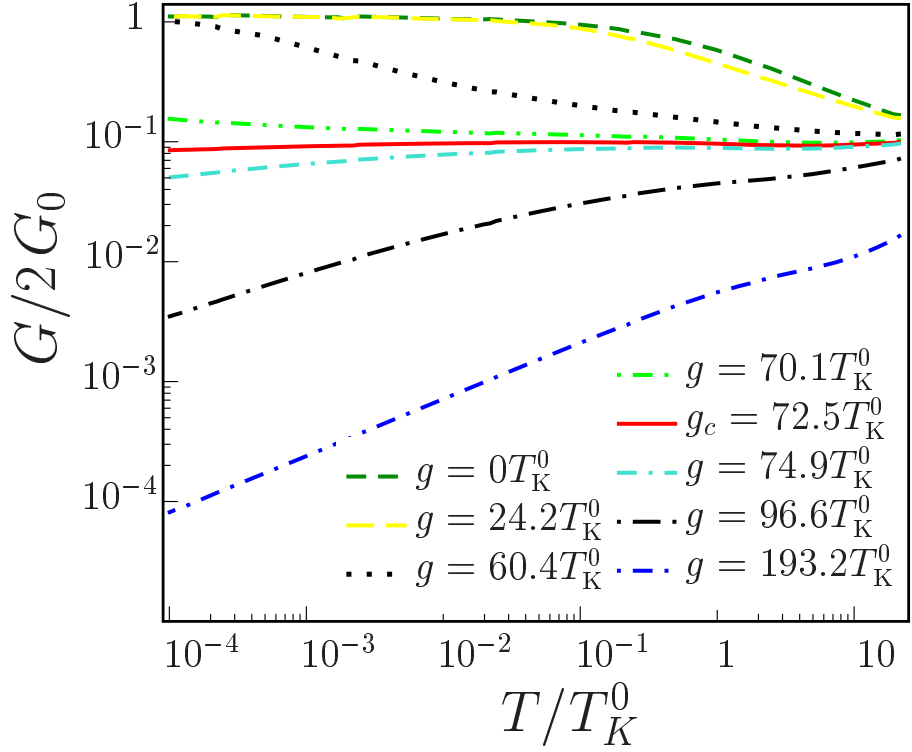}}
\caption{DC conductance. At $g<g_c$, the Kondo assisted 
conductance increases as temperature is lowered.
At the QCP ($g=g_c$) and inside 
the critical local-moment phase ($g>g_c$), 
the temperature dependence
has the forms of 
Eqs.~(\ref{conductance-dc-qcp},\ref{conductance-dc-bosonic}),
respectively. $G_0$ is $e^2/h$.
} \label{FIG3}
\end{figure}

The linear-response DC conductance ($G$) directly manifests the
critical nature of the Kondo effect. Its temperature 
dependence is 
given in Fig.~\ref{FIG3}.
On the Kondo side ($g<g_c$),
the conductance increases as the temperature is lowered,
reflecting the development of a Kondo resonance~\cite{Glazman.88,Ng.88}.
Inside the critical local-moment phase ($g>g_c$), the conductance vanishes
in a power-law form as temperature goes to zero. It vanishes at
$T=0$ since Kondo screening has been completely destroyed. At finite
temperatures, charge transport can still occur because of the 
finite effective
Kondo coupling at non-zero energies. 
Hence, the way the conductance goes to zero
characterizes the critical behavior of the Kondo coupling at the 
critical local-moment fixed point.
At the QCP ($g=g_c$),
the conductance behaves in a way intermediate between the two sides.
The zero-temperature conductance is finite, but is reduced from 
its counterpart on the Kondo side. The temperature-dependent 
part has a power-law form that is distinct from both the local-moment
and Kondo sides, characterizing the critical exponents unique 
to the QCP.

\begin{figure}[h!]
\centerline{\includegraphics[width=0.8\linewidth]{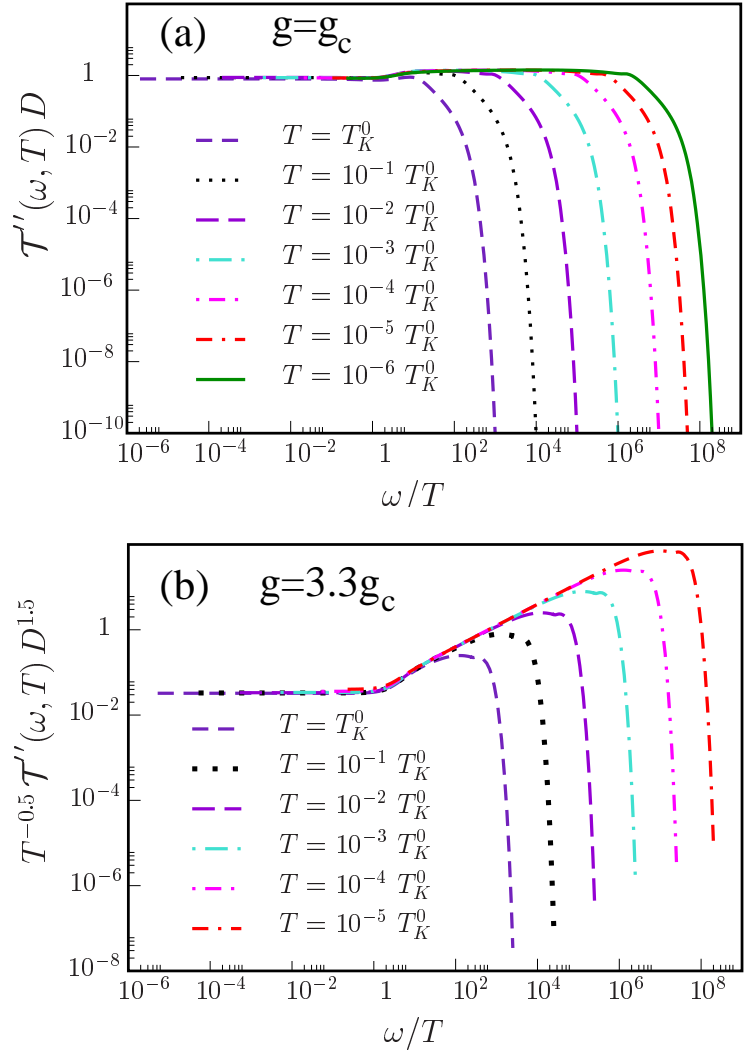}}
\caption{$\omega/T$ scaling. 
{\bf a},~The T-matrix as a function of $\omega/T$, at the 
QCP. For each temperature, the T-matrix falls on the universal
scaling curve until
$\omega$ reaches the order of
$T_K^0$. Notice that
${\cal T}''(\omega \rightarrow 0, T\rightarrow 0, 
\omega/T \rightarrow 0)$ is different from
${\cal T}''(\omega \rightarrow 0, T\rightarrow 0, 
\omega/T \rightarrow \infty)$.
{\bf b},~Similar result inside the critical local-moment phase,
at a $g>g_c$; the critical exponent is $0.5$.
} \label{FIG4}
\end{figure}

To understand the detailed temperature dependences,
we have analytically determined  the zero-temperature
T-matrix. At the QCP ($g=g_c$), its imaginary part is
${\cal T}''(\omega+i0^+,T=0)  = a + b |\omega|^{1/4}$.
On the local-moment side ($g>g_c$), the asymptotic
low-energy behavior is 
${\cal T}''(\omega+i0^+,T=0)  = c |\omega|^{1/2}$.
The exponents here reflect the critical exponents of the 
QCP and the critical local moment phase
with a dissipative spectrum given in Eq.~(\ref{dispersion}).
The results suggest that the linear response DC conductance 
is, at the QCP,
\begin{eqnarray}
G(T) = A + B T^{1/4} ,
\label{conductance-dc-qcp}
\end{eqnarray}
and, inside the critical local-moment phase ($g>g_c$),
\begin{eqnarray}
G(T) = C T^{1/2} .
\label{conductance-dc-bosonic}
\end{eqnarray}
Indeed, these fractional-power-law forms fit the numerical 
results shown in Fig.~\ref{FIG3}.
The success of this fitting suggests that the T-matrix
satisfies an $\omega/T$ scaling. This is verified by the scaling 
plots shown 
in Fig.~\ref{FIG4}a for $g=g_c$ and
in Fig.~\ref{FIG4}b for $g>g_c$.

AC transport and noise properties provide the means to probe 
the $\omega/T$ scaling. At the QCP,
the real part of the AC conductance is
\begin{eqnarray}
G'(\omega,T) = {\cal G}_{c0}(\omega/T) +
\omega^{1/4} {\cal G}_{c1}(\omega/T) ,
\label{conductance-ac-qcp}
\end{eqnarray}
whereas, inside the local-moment phase ($g>g_c$),
\begin{eqnarray}
G'(\omega,T)
 = \omega^{1/2} {\cal G}_b(\omega/T).
\label{conductance-ac-bosonic}
\end{eqnarray}
Here, ${\cal G}_{c0}$, ${\cal G}_{c1}$ and
${\cal G}_b$ are scaling functions.
The thermal current fluctuations (Johnson noise) 
spectrum, $S(\omega)$, is simply 
Eqs. (\ref{conductance-ac-qcp},\ref{conductance-ac-bosonic})
multiplied by $2\hbar \omega [1+n_B(\omega)]$, where the Bose factor
$n_B(\omega) \equiv 1/[{\rm e}^{\hbar \omega /k_B T} - 1]$.
$S(\omega)/\omega$, too, satisfies $\omega/T$ scaling at the QCP
and inside the critical local-moment phase.

We turn next to the feasibility of the experimental measurements.
The non-Fermi liquid behavior we have derived applies to the
scaling regime, the region in which universal properties arise.
We have so far assumed a spin-wave spectrum that is gapless.
Real ferromagnets contain interactions (such as spin-orbit
coupling) that break 
spin-rotational symmetry,
which will cut off the scaling regime at low energies.
The bulk spin-anisotropic interaction results in
a spin wave gap of about $0.25 ~{\rm K}$ for Ni
films~\cite{vanKampen.02}
and even smaller values in the case of permalloys.
The surface spin-anisotropic interactions yield a finite de-pinning energy
which, however, is negligibly small (of the sub mK 
range; see
the Formal Details section).
%% Methods section).
The upper cutoff energy of the scaling regime is given
by the Kondo scale, $T_K^0$, which, for single-molecular
transistors, can be of the order of $50-100$~K
(Refs.~\cite{YuN04_b,YuN04c}).
Hence, we can expect more than two decades of both
temperature and frequency over which the non-Fermi liquid
behavior occurs.
To study the AC noise 
spectrum, the frequencies needed fall in the range of about
a few gigahertz to a terahertz. 
Much of this frequency range is already
feasible experimentally: noise measurements up to 90 
gigahertz have been reported~\cite{Deblock.03}.

We have also assumed symmetric couplings of the 
dot to
the two leads, which ensure a zero local effective magnetic field 
for the local moment due to exchange ($h_{\rm loc}=0$). 
[The fringing fields from the ferromagnetic leads are expected 
to be small ($\ll 0.6$K for Ni)~\cite{Pasupathy.04}.]
Significantly different left/right couplings
will yield
a finite $h_{\rm loc}$ (which,
on the Kondo side, in turn results in a splitting 
of the Kondo resonances even in the AP 
configuration~\cite{Pasupathy.04,MartinekSBBKSv03,Choi.04}).
Some of the 
single-molecule transistors
reported
in Ref.~\cite{Pasupathy.04} indeed have approximately
symmetric couplings, as evidenced by the smallness of the splitting 
in the Kondo peaks. 
Even if it is nonzero, $h_{\rm loc}$ can be compensated by an
external local magnetic field~\cite{MartinekSBBKSv03}.
The latter can be generated by dc currents through nearby
conductors, as in magnetic random access memory
applications~\cite{Engel.02}.

So far, we have considered 
ferromagnetic metallic leads.
A variation is to use metallic leads
which are paramagnetic, but are nearly ferromagnetic.
The absence of static ordering implies that the local
effective field (from dot-lead exchange) 
vanishes under all circumstances. 
Spin waves are no longer sharply defined;
they do, nonetheless, appear in the form
of paramagnons, which provide a dissipative bosonic bath
with a fluctuation spectrum similar to that of Eq.~(\ref{dispersion}),
but with an exponent $1/3$, instead of $1/2$. 
This is still sub-ohmic, so with appropriate modifications to
the fractional exponent and scaling functions, essentially all
our results will still apply. 
(The almost magnetic nature of the leads here can generate multi-channels
of effective conduction electrons~\cite{Maebashi.04,Larkin.72}, 
which can influence the finite temperature properties on the 
Kondo side. This effect, however, is not expected to 
modify the quantum critical behavior.)
For instance, the conductance exponent
at $g=g_c$ is now $1/3$ (instead of $1/4$), and that for
$g> g_c$ becomes $2/3$ (instead of $1/2$).
Palladium would be one candidate material in this category.

Our results are insensitive to the presence or absence of particle-hole 
symmetry. In addition, there is no need to achieve multiple channels
of fermionic couplings. These make the
non-Fermi liquid behavior of
a ferromagnetic single-electron transistor
more robust than
that of either the multi-channel or multi-impurity Kondo 
systems. There is a simple reason behind this robustness. 
The phases of the system we have discussed -- the Kondo phase 
on the one hand and the critical local-moment phase on the other --
are genuinely distinct and must be separated by a quantum phase transition.
Since the transition is second order, 
a distinctive QCP must arise and quantum criticality in turn makes 
the system a non-Fermi liquid.
In the multi-channel or multi-impurity Kondo
systems~\cite{Varma.02},
however, all the stable phases are Fermi
liquids, making special symmetries -- channel degeneracy or particle-hole
symmetry -- necessary to reach any non-Fermi liquid state.

It is instructive to discuss the broader implications 
of our results.
Orthodox theory of any QCP~\cite{Sachdev,Hertz}
describes it in terms of order-parameter
fluctuations and maps it to a classical counterpart 
in elevated dimensions.
For the Bose-Fermi Kondo model
with a bosonic fluctuation spectrum given by Eq.~(\ref{dispersion}),
the corresponding
critical point is non-interacting~\cite{Zhu.04,Vojta.05,Glossop.05}.
On the other hand, 
the $\omega/T$ scaling implies~\cite{Sachdev}
that the critical point we are dealing with is a fully interacting
one. A further support for the interacting nature of the QCP
comes from Fig.~\ref{FIG4}a. It is seen that the T-matrix at 
the limit of zero temperature and $\omega \rightarrow 0$ 
[corresponding to $\omega/T \gg 1$ and yielding 
${\cal G}_{c0}(\infty)$ -- {\it cf.} Eq.~(\ref{conductance-ac-qcp})]
is different from that at the limit of 
$\omega=0$ and $T \rightarrow 0$ 
%%limit 
[so that $\omega/T \ll 1$,
giving rise to ${\cal G}_{c0}(0)$].
This is in contrast to non-interacting critical points of quantum
impurity models in which
the two limits are believed to be interchangeable~\cite{DamleSachdev.97}.
An experimental observation of either $\omega/T$ scaling, or
${\cal G}_{c0}(\infty) \neq {\cal G}_{c0}(0)$, 
would provide 
a direct signature of the inherent quantum nature of the QCP.

In addition to their intrinsic interest, 
our results will establish how the Kondo effect influences 
quantum criticality. 
The latter not only is directly relevant to the quantum-critical
heavy fermions~\cite{Coleman.05,Si.01,Paschen.04},
whose properties strongly deviate from
the expectations of the order-parameter-fluctuation
picture,
but also illustrates the central role of
quantum entanglement effects -- beyond the
fluctuations of order parameter -- in
quantum criticality in general.

%%\\[2ex]
\vspace*{0.4 cm}
\noindent{\bf \Large Conclusions}
%%\vspace*{0.2 cm}

To summarize, we propose the 
ferromagnetic single-electron transistor
as a model
system to study non-Fermi liquid states and quantum criticality.
We have shown that the quantum critical point is robust
theoretically and feasible in experimental implementation.
Finally, our results imply that
the system, readily measurable beyond the equilibrium
regime, will serve as an ideal setting for the much-needed
explorations of the non-equilibrium properties at quantum
criticality. All these features make it a tunable spintronic
system to study issues that are important to a broad array
of strongly correlated electron materials.

%%\\[2ex]
\vspace*{0.4 cm}
%%\noindent{\bf \Large Methods}\\
\noindent{\bf \Large Formal Details}\\
%%\vspace*{0.1 cm}
%%\vspace*{0.2 cm}
\noindent{\bf Derivation of the Bose-Fermi Kondo model.}
%%\\
Consider the regime where the dot has 
an odd number of electrons and the valence-fluctuation
energy $\Delta$ is large compared to the resonance width $\Gamma$. 
At temperatures well below $\Delta/k_B$,
valence fluctuations can occur only virtually
and the dot is effectively a localized  moment.
Hybridization processes between the dot and 
electrodes
give rise
to an antiferromagnetic coupling between this moment and
the spins of the low-energy conduction electrons in the lead
metal. The conduction-electron spin in a direction perpendicular 
to the ordered moments of the ferromagnets is a linear combination
of the transverse part of the triplet quasiparticle-hole excitations
and the spin waves. 
So we expect the local moment to be coupled not only
to the spins of the quasiparticles but also to
the spin waves.
In order to determine the specific value of the bosonic
coupling constant, we 
go beyond these general arguments and 
carry out a generalized Schrieffer-Wolff transformation
followed by a projection (${\cal P}$) onto the local-moment subspace:
\begin{equation}
\label{eq:effH}
{\mathcal H}_{\mbox{bfk}}
\,={\cal P}\,e^{S}\,(H_{\mbox{\small{0}}}+
H_{\mbox{\small{lead}}}+H_{\mbox{\small{hyb}}})\,e^{-S}{\cal P},
\end{equation}
where $H_{\mbox{\small{0}}}$ is the charge-conserving part of the dot
Hamiltonian.
The generator of the canonical transformation 
($S$) follows from the requirement
that
$[H_{\mbox{\small{0}}}+H_{\mbox{\small{lead}}},S]\,=\, 
H_{\mbox{\small{hyb}}}$, and can be formally expressed as \cite{Schork.94},
\begin{equation}
\label{eq:liouville}
S\,=\, (L_{\mbox{\small lead}}+L_{{\mbox{\small 0}}})^{-1}\,H_{\mbox{\small hyb}} ,
\end{equation}
where the Liouville operators are defined as
$L_{x}\, A\,=\, [H_{x},A]$.
We model the ferromagnetic metal of a lead in terms of 
a one-band Hamiltonian,
$H_{\mbox{\small lead}}\,=\,
\sum_i H_{\rm kin,i} 
-\frac{2\,u}{3}\sum_{{\bf r},i} \bigl(
{\bf s}_i
(
{\bf r}
)\bigr)^2$,
where $H_{\rm kin,i} =
\sum_{\bf k,\sigma} \epsilon_{{\bf k}} c_{{\bf k} \sigma i}^{\dagger}
c_{{\bf k}\sigma i}$,
with $i=L,R$ being the lead index. 
The interaction term is decoupled in terms of the vector boson field
$\vec{\phi}$ as follows,
\begin{eqnarray}
\label{lead_ham}
H_{\mbox{\small lead}}&=&
\sum_i H_{\mbox{\small   kin},i} 
+ \sum_{
{\bf r},i
} 
\bigg [ \frac{1}{2} u 
\vec{\phi}_i^{\dagger}(
{\bf r}
)
\cdot
\vec{\phi}_i^{}(
{\bf r}
) \\
&+& \lambda
[\phi_i(
{\bf r}
)+\vec{\phi}_i^{\dagger}(
{\bf r}
)]\cdot 
{\bf s}_i
({\bf r})
\bigg ] \nonumber ,
\end{eqnarray}
where $\lambda = \sqrt{2/3}u$.
Within the ferromagnetic-metal phase, the $z$-component ${\phi}_{z,i}$
acquires a static value, $m_i$, and the transverse components 
${\phi}_{\beta,i}$, $\beta = x,y$, describe the spin waves.
Inserting 
Eqs.~(\ref{eq:liouville},\ref{lead_ham})
in Eq.~(\ref{eq:effH}) 
leads to a Bose-Fermi Kondo model,
\begin{eqnarray}
{\mathcal H}_{\mbox{bfk}}&=&
J \sum_{i}{\bf S} \cdot {\bf s}_{i} +
\sum_{{\bf k},i,\sigma} 
\tilde{\epsilon}_{{\bf k}\sigma i}~
c_{{\bf k}\sigma i}^{\dagger} c_{{\bf k} \sigma i} + g S_{z} \sum_i m_i 
\nonumber\\
&+&
g S_{\beta} 
\sum_{\beta,{\bf q},i}
(\phi_{\beta,{\bf q},i} +
\phi^{\dagger}_{\beta,{\bf q},i} )
+ \sum_{\beta,{\bf q},i}
\omega_{\bf q}\,
\phi_{\beta,{\bf q},i}^{\;\dagger} \phi_{\beta,{\bf q},i} ,
\label{hamiltonian-bfk-n=2}
\end{eqnarray}
with the coupling constant $g \sim u t^2/\Delta^2_{}$ which
gives rise to the expression for $g$ 
stated earlier 
once we recognize
that, in itinerant ferromagnets, $\rho_0 u \sim 1$.
The conduction electron dispersion in each lead is Zeeman split:
$\tilde{\epsilon}_{{\bf k}\sigma i} = \epsilon_{\bf k} + \lambda m_i 
\sigma$. However, only one linear combination of the conduction electrons
from the two leads is coupled to the dot spin~\cite{Glazman.88,Ng.88}:
$c_{{\bf k}\sigma } 
=( c_{{\bf k}\sigma L } 
+c_{{\bf k}\sigma R} )/\sqrt{2}$.
The energy dispersion for this linear combination is
spin-independent, and will be denoted 
as $E_{\bf k}$.
%%S
A similar linear combination arises for the spin waves from the two leads:
$\Phi = (\phi_L+\phi_R)/\sqrt{2}$.
The local effective field, $h_{\rm loc} = g \sum_i m_i$, vanishes
in the AP case ($m_L=-m_R$).
The Hamiltonian (\ref{hamiltonian-bfk-n=2})
has an easy-plane anisotropy.
Taking advantage of the observation~\cite{Kirchner.05,Zhu.04}
that the scaling
properties of both the quantum critical point and the 
critical local-moment phase in this case are similar to those of its
SU(2) and SU(N) counterparts,
we will generalize the local moment to 
a form with SU(N) symmetry, and the conduction electrons to a form
with SU(N)$\times$SU(N/2) 
symmetry~\cite{Parcollet.98,Cox.93}.
After a rescaling of the Kondo
coupling to order $1/N$, and the coupling to the spin waves to order
$1/\sqrt{N}$, we arrive at the Hamiltonian:
\begin{eqnarray}
{\mathcal H}_{\mbox{\tiny BFK}}
&=&
({J}/{N}) \sum_{\alpha}{\bf S} \cdot {\bf s}_{\alpha} +
\sum_{{\bf k},\alpha,\sigma} E_{\bf k}~c_{{\bf k} \alpha \sigma}^{\dagger} 
c_{{\bf k} \alpha \sigma} \nonumber \\
&+&
({g}/{\sqrt{N}})
{\bf S} \cdot {\bf \Phi} + \sum_{\bf q} \omega_{\bf q}\,{\bf
\Phi}_{\bf q}^{\;\dagger}\cdot {\bf \Phi}_{\bf q}. \label{hamiltonian-bfk}
\end{eqnarray}
Here, ${\bf S}$, $c_{{\bf k} \alpha \sigma}$, and 
${\bf \Phi}_{\bf q}$ 
denote the local moment,
fermionic bath and bosonic bath, respectively, 
${\bf \Phi} =
\sum_{\bf q} ({\bf \Phi}_{\bf q} + {\bf \Phi}_{\bf q}^{\dagger})$
and 
${\bf s}_{\alpha} = 
(1/2) \sum_{{\bf k} \sigma,{\bf k}^{'}\sigma^{'}}
c_{{\bf k} \alpha \sigma}^{\dagger}
\vec{\tau}_{\sigma,\sigma^{'}}
c_{{\bf k}' \alpha \sigma'}$
where $\vec{\tau}$ is the 
vector of 
Pauli matrices.
$\sigma=1,...,N$ labels spin indices and $\alpha=1,...,N/2$ 
the channel indices for conduction electrons.

The spin-wave spectrum determines the dissipative bosonic bath spectrum.
We end up with Eq.~(\ref{dispersion})
using 
the spin-wave dispersion $\omega_{\bf q} = \rho_s q^2$,
where $\rho_s$ is the spin-wave stiffness.

%%\\
\noindent{\bf Dynamical large-N method.}
%%\\
The dynamical saddle-point equations for the large-$N$
limit of the Bose-Fermi Kondo Hamiltonian
have been 
described in Ref.
\cite{Zhu.04}.
In addition to allowing a controlled solution in general,
this limit
allows us to 
calculate
the conduction electron T-matrix.
The saddle-point equations are 
expressed
in terms of
$G_f(\tau)$ and $G_B(\tau)$,
the propagators for the slave-$f$-fermion of the dot spin
and for the $B$-field representing the $f^{\dagger}c$
combination~\cite{Zhu.04,Parcollet.98,Cox.93},
respectively.
Once these quantities are 
determined, the T-matrix is given
by the Fourier transform
of ${\cal T}(\tau) = -\frac{1}{N} G_B(-\tau)G_f(\tau)$.
The large N limit for the Kondo phase ($g<g_c$) is
described by the fixed point at $g=0$, which is a
multi-channel fixed-point~\cite{Parcollet.98,Cox.93}.
For the critical local moment phase ($g>g_c$) and the QCP
($g=g_c$), on the other hand, both the large N
and the $N=2$ cases describe non-Fermi liquid states
and the two cases have similar critical
behaviors~\cite{Kirchner.05,Zhu.04}.

At $T=0$, the large-$N$ limit is solved analytically,
by making the scaling ansatz 
for both the leading and sub-leading terms,
\begin{equation}
G_f(\tau)= \frac{A_1}{\tau^{\alpha_1}}
+\frac{A_2}{\tau^{\alpha_2}}
; ~~~ G_B(\tau)=\frac{B_1}{\tau^{\beta_1}}
+\frac{B_2}{\tau^{\beta_2}} ,
\label{gf-leading}
\end{equation}
and inserting these into the saddle-point equations.
Consider first the leading order. 
For the critical local-moment phase,
$\alpha_1=1/4$ and $\beta_1=5/4$.
As a result, ${\cal T}(\tau) \sim 1/\tau^{3/2}$, corresponding to
an $\omega^{1/2}$ dependence.
[In the more general case with $\omega^{1-\epsilon}$ replacing
$\sqrt{\omega}$ in Eq.~(\ref{dispersion}), the result
is $\omega^{\epsilon}$.]
For the QCP ($g=g_c$), $\alpha_1=1/4$ and $\beta_1=3/4$, yielding 
${\cal T}(\tau) \sim 1/\tau$, corresponding to a constant 
$ {\cal T}''(\omega+i0^+)$
[the critical exponent at the leading
order (being equal to 1), though neither the critical amplitude
(the constant $a$ given below) nor 
the finite-temperature behavior 
(Fig.~\ref{FIG4}a), 
%%at the same order
was anticipated by the general
argument of Ref.~\cite{Vojta.03}].
The $\omega$-dependence comes 
from the subleading terms of $G_f(\tau)$ and $G_B(\tau)$.
We find $\alpha_2=1/2$ 
and $\beta_2=1$, making the dominant subleading contribution
to ${\cal T}(\tau)$ to be $1/\tau^{5/4}$ which, in turn, corresponds
to an $\omega^{1/4}$ dependence.
(For general $\epsilon$, this becomes $\omega^{\epsilon/2}$.)
The final results,
for $g=g_c$ and $g>g_c$,
are given in the Results and Discussion section, just before 
Eq.~(\ref{conductance-dc-qcp}),
with the numerical factors $a=\tan(\pi/8)/(8 N \rho_0)$,
$b=\pi\bigl[5\tan(\pi/8)\bigr]^{1/2}/\bigl[N \rho_0K_0g_c
\Gamma^3(1/4)\bigr]$,
and $c=(1/8)\Gamma(5/4)\bigl[3\tan(\pi/8)/2\bigr]^{1/2} 
J/[N K_0g\Lambda^{1/4}]$.
Here,
$\Gamma(x)$ is the gamma function
and 
$K_0=[\Gamma (3/2)/4\pi^2\rho_s^{3/2}]^{1/2}$ is related to 
the proportionality factor on the right hand side
of Eq.~(\ref{dispersion}).

At finite temperatures, the saddle-point
equations
in the large-$N$ limit
are solved numerically
in terms of real frequencies~\cite{Zhu.04}.
To check the robustness of our results with respect
to particle-hole symmetry breaking, we have also
solved the extension of the above
saddle-point equations to the Bose-Fermi Anderson model in the 
local-moment regime without particle-hole
symmetry (setting $N=2$):
\begin{eqnarray}
H_{\mbox{
bfam}}&=& \sum_{{\bf k},\sigma} E_{\bf k}~c_{{\bf k}
\sigma}^{\dagger}
c_{{\bf k} \sigma}+ \varepsilon_d  \sum_{\sigma} 
d^{\dagger}_{\sigma}d^{}_{\sigma}
+ U n_{d\uparrow} n_{d\downarrow} \nonumber  \\
&+&
 t \sum_{{\bf k},\sigma} \biggl (c_{{\bf k}
  \sigma}^{\dagger} d^{}_{\sigma} + \mbox{h.c.} \biggr)
 + g 
{\bf S}_d \cdot {\bf \Phi} \nonumber \\
&+&
\sum_{\bf q} \omega_{\bf q}\,{\bf
\Phi}_{\bf q}^{\;\dagger}\cdot {\bf \Phi}_{\bf q}
,
\label{bfam}
\end{eqnarray}
with $ {\bf S}_d=(1/2) \sum_{\sigma,\sigma^{'}} d^{\dagger}_{\sigma}
\vec{\tau}_{\sigma,\sigma^{'}} d^{}_{\sigma^{'}}$, and 
$\varepsilon_d  \neq - U/2$.
The scaling behavior is found to be the same as that of either 
the numerical results in the large N limit we have considered,
which has a particle-hole symmetry, 
or (where a comparison can be made) the  analytical results 
in the same limit.

\noindent{\bf Calculation of the transport quantities.}
%%\\
The linear-response DC conductance is
\begin{eqnarray}
G &=& \lim_{V\to0} \frac{dI}{dV} \nonumber \\
&=& \frac{e^2}{{\hbar}} 
N\Gamma \int_{-\infty}^{\infty}\,d\epsilon\, 
\frac{1}{\pi} {\cal T}''(\epsilon+i0^+)
\biggl[ -
\frac{\partial f(\epsilon)}{\partial \epsilon} \biggr],
\label{dc-conductance}
\end{eqnarray}
where
$f(\epsilon)$ is the Fermi-Dirac distribution function. 
Likewise, the real part of the linear-response AC conductance 
(the current response to an oscillating bias voltage)
can also be related to the scattering 
T-matrix~\cite{Sindel.04,Buettiker-prl.93}
under the condition of vanishing charge accumulation:
\begin{eqnarray}
G' (\omega) &=& \frac{e^2}{{2\hbar}} 
N\Gamma \int_{-\infty}^{\infty}\,d\epsilon\, 
\frac{1}{\pi} {\cal T}''(\epsilon+i0^+) \times \nonumber \\
&& \hspace*{0.3 cm} 
\biggl[ f(\epsilon - \hbar \omega) - 
f(\epsilon + \hbar \omega) \biggr] / (\hbar \omega) .
\label{ac-conductance}
\end{eqnarray}
The thermal current-fluctuation noise spectrum is simply related 
to $G' (\omega)$ through the fluctuation-dissipation theorem.
The result is given in 
Eqs.~(\ref{conductance-ac-qcp},\ref{conductance-ac-bosonic}),
with the following limiting values:
${\cal G}_{c0} (x\rightarrow \infty) = 2 a G_0 \Gamma / \hbar$,
${\cal G}_{c1} (x\rightarrow \infty) = 8b G_0 \Gamma / (5 \hbar)$,
and ${\cal G}_{b} (x\rightarrow \infty) = 8c G_0 \Gamma / (5 \hbar)$.

%%\vskip 0.5 cm
\noindent{\bf De-pinning energy of spin waves at the surface.}
%%\\
The form of the coupling between the local moment and spin waves,
given in the third term on the right hand side of 
Eq.~(\ref{hamiltonian-bfk-n=2}), has been derived in the absence 
of any surface spin-anisotropic interaction. The latter introduces
a pinning of the spin waves of sufficiently long
wavelengths~\cite{Kittel.58}.
Below the pinning wavevector, $q<q_{\rm pin}$, the wavefunction
at the surface is linearly~\cite{Kittel.58,Pincus.60} proportional to $q$;
in turn, the coupling constant between $S_{\beta}$ and
$\phi_{\beta,{\bf q},i}$ is proportional to $q$, changing
the spectrum of the dissipative bath from Eq.~(\ref{dispersion})
to $\sim \omega^{3/2}$. From the known result~\cite{Kittel.58,Pincus.60}
$q_{\rm pin} a \approx E_{\rm sa}/E_{\rm ex}$, 
where $a$ is the lattice constant and 
$E_{\rm sa}$ and $E_{\rm ex}$ are respectively the surface anisotropic
and bulk isotropic exchange energies,
we can estimate the corresponding pinning energy scale
$E_{\rm pin} \approx E_{\rm ex} (q_{\rm pin} a)^2 = 
E_{\rm sa}^2/E_{\rm ex}$. 
(Here, we have used the mean-field expression
$\rho_s \approx E_{\rm ex} a^2$.)
Being smaller than $E_{\rm sa}$ by a factor
$E_{\rm sa}/E_{\rm ex}$, the pinning energy is expected to be rather small.
For Fe-Co films, for instance, $E_{\rm sa} \sim 1$~K
(corresponding to a typical surface anisotropy~\cite{Schreiber.96}
of the order $0.1~{\rm erg/cm^2}$)
and $E_{\rm ex} > \sim 1000$~K
(based on the Curie temperature),
giving rise to an $E_{\rm pin} < 1$~mK.

\noindent{\bf Acknowledgments}
We would like to thank E. Abrahams, K. Ingersent, C. M. Varma, and 
G. Zar\'{a}nd for useful discussions. This work has 
been supported by DFG (SK), Robert A. Welch Foundation,
NSF Grant No.\ DMR-0424125 (SK, LZ, \& QS), 
the Alfred P. Sloan Foundation,
the David and Lucille Packard Foundation
and NSF Grant No.\ DMR-0347253 (DN).

%%%%%%%%%%%%%%%%%%%%%%%%%%%%%%%%%%%%%%%%%%%%%%%%%%%%%%%%%%%%%%

%%%%%%%%%%%%%%%%%%%%%%%Figure Captions%%%%%%%%%%%%%%%%%%%%%%%%%%%%%%%%%%%%%%%%
%%%%%%%%

\end{document}